# Gain limits of a Thick GEM in high-purity Ne, Ar and Xe

**J. Miyamoto[a], A. Breskin[a,1] and V. Peskov[a,b]**

[a] Department of Astrophysics and Particle Physics
Weizmann Institute of Science,
76100 Rehovot, Israel
[b] CERN,
1206 Geneva, Switzerland

E-mail: Amos.Breskin@weizmann.ac.il

ABSTRACT: The dependence of the avalanche charge gain in Thick Gas Electron Multipliers (THGEM) on the purity of Ne, Ar and Xe filling gases was investigated. The gain, measured with alpha-particles in standard conditions (atmospheric pressure, room temperature), was found to considerably drop in gases purified by non-evaporable getters. On the other hand, small $N_2$ admixtures to noble gases resulted in high reachable gains. The results are of general relevance in the operation of gas-avalanche detectors in noble gases, particularly that of two-phase cryogenic detectors for rare events.

---

[1] Corresponding author

# Contents

## 1. Introduction

Gas-avalanche detectors operating in noble gases have raised considerable interest. They have been extensively studied at atmospheric and high pressures, for a variety of applications [1-4], mostly in Ar, Xe as well as in Kr, Ne and He [5-13]. Of particular interest are Cryogenic Two-Phase Detectors [14,15] with scintillation and/or charge multiplication in the saturated vapor above a noble liquid, for the detection of rare-events (e.g. Dark-Matter searches and neutrino physics) [3-4]. The operation of Gas Electron Multipliers (GEM) in noble gasses, Ne, Ar, Xe and Kr without photon-quenching admixtures, resulted in relatively high gains; e.g. a gain of 700 was obtained with a single-GEM element in Ar [16] and a gain of $10^4$ was obtained in Ar with a triple-GEM [17]. It was suggested that these relatively high gains could be a result of avalanche confinement within the small (60 micron diameter) GEM holes, preventing instabilities initiated by secondary photon-feedback effects [16,17]. More recent studies indicated that these high gains could originate from Penning effects due to gas impurities (e.g. materials out-gassing) [18, 19]; indeed, in Ne and He under conditions where the impurities were removed (frozen at low temperatures), the GEM's maximum achievable gains $A_{mGEM}$ were close to unity [18, 19]. A similar effect of gain drop with gas purity was observed earlier in single-wire detectors operating in pure noble gases (impurity concentration $< 10^{-5}$ %) [5,6]. In the same works, it was found that the higher the ionization potential of the noble gas is, the lower is the maximum gain.

In recent years, some groups reported that Thick GEMs (THGEMs [20]) can be also used in noble-liquid TPCs and Cryogenic Two-Phase noble-liquid detectors; in some conditions THGEMs yielded even higher gas gain than GEMs [21-25]. All these early measurements have been performed without particular purification of the noble gases and without any quantitative control or monitoring of gas impurities. In a recent work, however, the same tendency was observed with a THGEM: in a getter-purified Ar the maximum achievable gain $A_{mTHGEM}$ dropped by a factor of ~10 compared to a non purified Ar circulating in the same experimental chamber [26].

All these observations indicate upon a general, similar, impurity-dependent trend, regardless of the type of the gaseous multiplier.

The aim of the present work was to perform systematic studies of the effect of impurities on the maximum achievable gain in THGEMs in Ne, Ar and Xe at normal conditions. Investigations in these gases are very relevant to the operation of current Two-Phase detector prototypes and to the design of large-volume detectors of rare events in which charge signals are measured by a multiplication process in addition to the primary scintillation event.

## 2. Methodology

The experiments were carried out in an ultra-high vacuum (UHV) system described in ref [26]. The setup (fig. 1) consists of two cascaded THGEMs made of FR-4 by a standard printed-circuit board production technique (mechanical drilling and etching of rims around the holes) [26]. The 30 x 30 mm$^2$ THGEM electrodes had the following geometry: thickness = 0.4 mm; hole diameter = 0.5 mm; hole spacing = 0.9 mm; rim size around the hole edge = 0.1mm. They were assembled with a drift mesh placed at 10 mm in front of the first THGEM; the second THGEM was followed by an anode placed at 1.6 mm from the electrode.

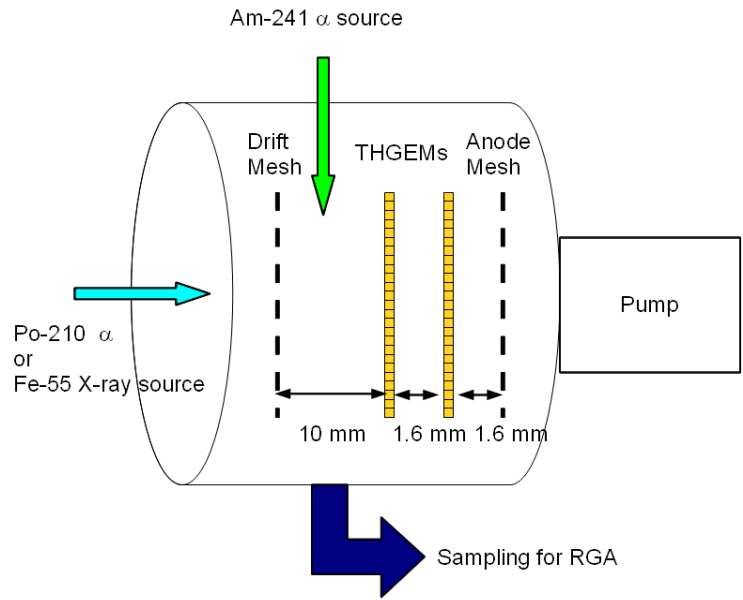

*Fig. 1. Schematic view of the experimental setup*

In the *flow mode*, the gas (or gas mixture) was continuously flowing from the bottle through the chamber, without purification, regulated by a mass-flow controller; in a *closed mode*, the chamber was closed and the gas was circulating through a hot (300-350 deg.C) non-evaporable getter (SAES-ST707) by convection, driven by the temperature gradient across the 10 cm-long getter housing. The gas composition (impurities levels) was constantly monitored with a Residual Gas Analyzer (Stanford Research Systems type 200 and MKS Microvision Plus), as described in [27]. Gases of the following initial purities were used: Ar = 99.999%; Ne = 99.996% and Xe = 99.999%.

The detector was operated either in single-THGEM or in double-THGEM mode; the avalanche electrons were recorded on the anode of the first or second THGEM, respectively. The detector was irradiated in different experiments either with $^{55}$Fe 5.9 keV x-rays or with alpha-particles from $^{241}$Am (foil-coated, ~ 4 MeV alphas) and $^{210}$Po (~ 4.3 MeV alphas) sources. X-rays irradiated the chamber, normal to the 10 mm drift gap, through a 50 micron thick Kapton window; due to electronic noise, x-ray induced charge signals were visible only at gains above few hundreds. The alpha- sources were mounted inside the chamber (Fig. 1) emitting alpha particles parallel (Ne and Ar studies) or perpendicular (Xe studies) to the THGEM electrodes; these sources were found very adequate for low-gain measurements, depositing sufficient ionization electrons to detect charges down to the ionization-collection mode (gain ~ 1). Besides gain investigations in *closed mode* (getter), comparative measurements were made in noble-gases in *flow mode*, with admixtures of $N_2$, to assess the role of the latter on the gain.

## 3. Results

### 3.1 Measurements in purified Ne

In these measurements, in *closed-mode,* the UHV chamber and the gas line were pumped down to high vacuum (~ $10^{-7}$ Torr) prior to Ne filling. The residual impurities introduced with the gas and others, resulting from outgasing from chamber and detector components, were gradually removed by gas circulation through the heated getter. The maximum gain was assessed while constantly monitoring the impurities level (nitrogen, water and oxygen) with the RGA.

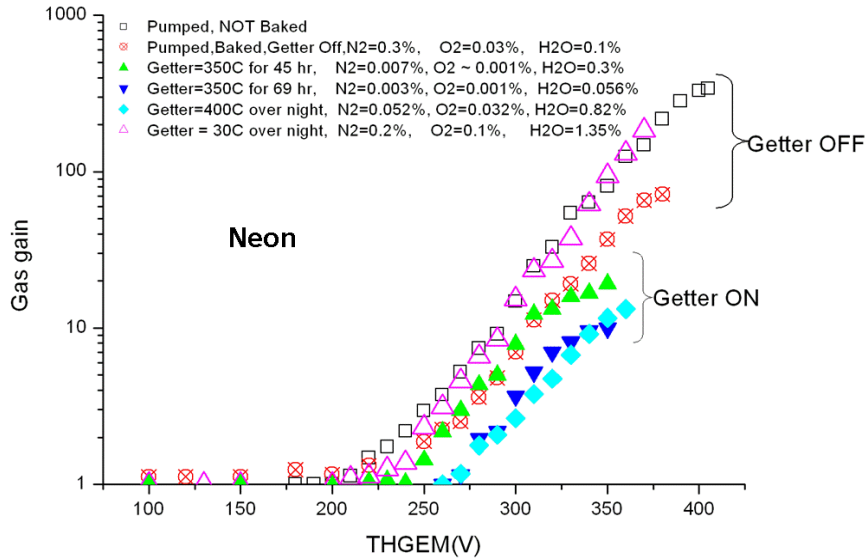

*Figure 2. Gain curves with a single-THGEM of Fig. 1, in a closed chamber filled with 1 atm Ne, irradiated with a $^{241}$Am alpha source. The closed symbols are data recorded with getter kept at 350 or 400ºC. The open symbols are for measurements without getter or with getter kept at low temperature (ineffective). In purified Ne, the maximum achievable gain $A_{mTHGEM}$ dropped by a factor 30-40.*

The gain curves recorded in Ne with a single-THGEM are plotted in Figure 2. With the activated getter and gradually reduced impurities, the maximum gain dropped by more than an order of magnitude. Note that the operation voltage increased (for a given gain value) after the respective drops of $N_2$ and $H_2O$ levels to 0.003% and 0.06% (the getter was kept at 350 ºC for about 3 days). A further increase in the getter's temperature, to 400 ºC, did not result in any noticeable changes in the gain; the impurities level slightly increased, possibly due to outgasing from the getter's housing.

In order to verify the getter effectiveness, its temperature was lowered to 30 ºC with the purified gas left in the chamber over night. The gain curve recorded at this getter temperature was very similar to the one recorded before turning on the getter (Figure 2); the impurities level ($N_2$, $O_2$, $H_2O$) rose significantly due to outgasing, as indicated in the figure.

In another measurement, the gain of a double-THGEM was measured in a similar environment (1 atm Ne in *closed mode* with a getter); the gain curves are plotted in Figure 3. A noticeable gain drop was already observed 3 hours after the getter was turned on; the gain kept dropping while the getter was active. After 50 hours of purification, the maximum attainable gain was as low as 30, only twice higher than with a single-THGEM (see Figure 2). Gains of 300 were reached in the double-THGEM, at best, with the getter turned off for 24 hours.

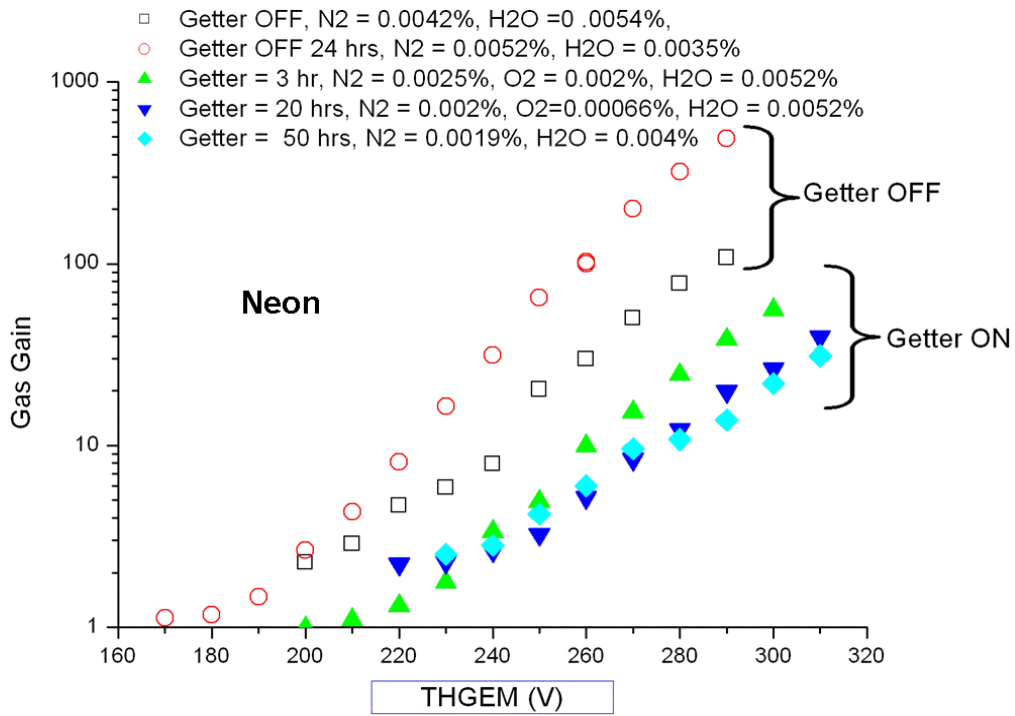

*Figure 3. Gain curves in the Double-THGEM of Fig. 1, in a closed chamber filled with 1 atm Ne, with a $^{241}$Am alpha-source. The closed symbols are data recorded with a getter kept at 350ºC. The open symbols are for measurements without getter or with getter kept at low temperature (ineffective). In purified Ne, the overall gain of the double-THGEM dropped by a factor of ~30.*

## 3.2. Measurements in purified Ar

A similar study in *closed mode* was done with a single-THGEM irradiated with a $^{241}$Am alpha source in a chamber filled with argon. The results are shown in Figure 4. The first measurement was made without turning on the getter; it showed a gain of ~ 100 with $N_2$ level ~ 0.27% and $H_2O$ ~ 1.1%. Three hours after the getter was turned on, the gain dropped by a factor of ~ 5, to a value of ~20; after a day of purification, it dropped to a level of ~ 3. The gain slightly dropped within the following two days, with no significant variations in the operation voltage. This is in contrast to the large shift in operation voltage observed above in Ne (Figures 2 and 3). It is clear that the mixtures Ne-$N_2$ and Ar-$N_2$ are fundamentally different. For example, there was no evidence in literature for Penning effect in Ar-$N_2$ indicating the lack of favorable conditions for energy transfer from excitation to ionization while it is known that Ne-$N_2$ forms a Penning mixture [28] (more details on Penning mixtures can be found in [29-32]). This point will be further discussed below.

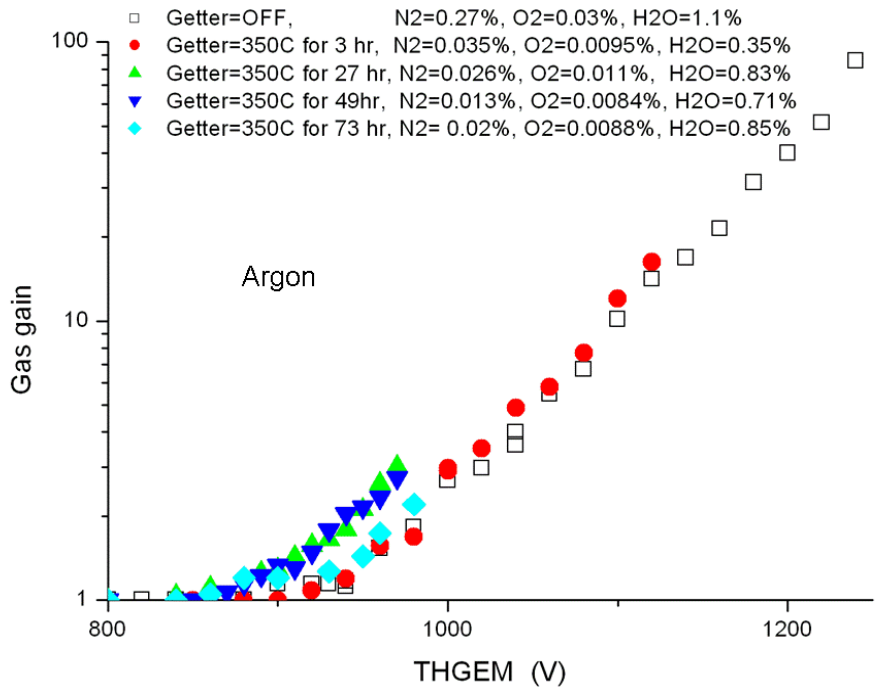

*Figure 4. Gain curves of a single-THGEM in 1 atm Ar, measured with a getter in a closed mode using the $^{241}$Am alpha-source. The open symbols correspond to data without the getter and the closed symbols are for data taken with the getter on. As in Ne, the maximum achievable gain dropped in pure Ar by a factor of ~30.*

## 3.3 Measurements in purified Xe

Gain curves measured in Xe with the $^{210}$Po alpha-source are plotted in Figure 5. Only a minor gain drop (roughly a factor of two) compared to the situation with Ne or Ar was observed when the getter was turned on. A possible explanation will be presented in section 4.3. Note, however, that this is a very preliminary result that requires further investigations because the Xe gas used here contained more impurities than Ar or Ne and the effective getter temperature, duration of purification, and other parameters have not been yet optimized.

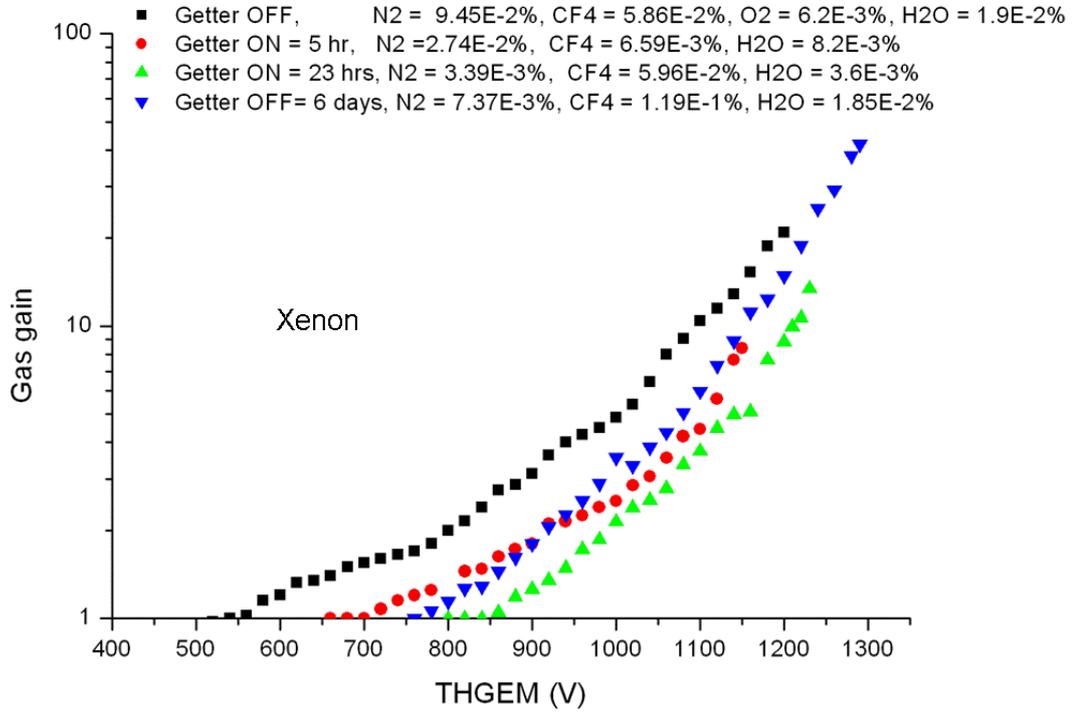

*Figure 5. Gain curves of a single-THGEM in 1 atm Xe, measured with a getter in a closed mode with the $^{210}$Po alpha-source. The open symbols correspond to data without the getter and the closed ones are for data taken with the getter on. The maximum achievable gain dropped in purified Xe by a factor of 2-3 only.*

## 4. Possible factors influencing the maximum gain in Ne, Ar and Xe

The main result obtained in this work is very clear: the cleaner the noble gas the lower is the maximum achievable gain of a THGEM.

In a single-THGEM operating in quenched gas mixtures, the maximum achievable gain is restricted by the Raether limit (for details see [33] and references therein):

$$A_m \, n_0 \approx 10^7 \text{ electrons} \qquad (1)$$

where $n_0$ is the number of primary electrons created in the drift region of the detector by ionizing radiation and $A_m$ is the maximum achievable gain above which breakdowns occur. As follow from formula (1), with alpha-particles inducing typically $n_0 \sim 10^5$ electrons, the maximum achievable gain would be $\sim 10^2$; this was indeed observed in this work at high concentration of impurities (note, that in multi-element electron multipliers, e.g. cascaded THGEM detectors, due to electron diffusion the Raether limit is higher, reaching values up to $A_m n_0 \approx 10^8$ electrons [33]).

In high-purity gases, however, the maximum achievable gain was as low as ~ 10, indicating that additionally, an onset of avalanche-feedback mechanisms restricts the maximum achievable gain. As was shown in [33], feedback-induced breakdown appears when one of the following conditions is satisfied:

$$Ab\gamma_{ph} = 1 \qquad (2)$$

or

$$Ac\gamma_{+} = 1 \qquad (3),$$

where b and c are coefficients describing the fraction of avalanche-induced photons and ions reaching the THGEM cathode at a given gain A; $\gamma_{ph}$ and $\gamma_{+}$ are the respective secondary-electron emission coefficients for photons and ions.

Regarding the photon-induced electron emission, it was shown in [33] that:

$$\gamma_{ph} = \int Q(V, E\nu) S(V, E\nu) \, dE\nu \qquad (4)$$

where $Q(V, E\nu)$ is the quantum efficiency of the cathode as a function of the voltage V applied to the detector and the photon energy $E\nu$; $S(V, E\nu)$ is the emission spectrum of the Townsend avalanches. Typically for metals $Q(V, E\nu)$ sharply (by orders of magnitude) increases with the photon energy; as a result, in most cases the vacuum ultraviolet (VUV) part of the avalanche emission spectrum has the strongest contribution to the value of $\gamma_{ph}$. In purified noble gases the avalanches strongly emit in the VUV (eximer spectra) [34], with $\gamma_{ph}$ depending on the gas: it will be the highest for He and Ne (emitting in the far VUV range: 60 -100 nm), lower for Ar (130 nm) and the lowest in in Xe (175nm).

As was shown in [5, 6], the probability to create an ion-induced secondary-electron is:

$$\gamma_{+} = k_g(V) (E_i - 2\varphi), \qquad (5)$$

where $k_g(V)$ is a coefficient (depending on the gas and the applied voltage), $E_i$ is the ionization potential of the gas and $\varphi$ is the work function of the cathode. As follows from the formula (5), $\gamma_{+}$ linearly increases with $E_i$; therefore it will be larger for He and Ne than, for example for Xe. In [5, 6] both ion- and photon-feedback effects were clearly observed in noble gases; it was proven that in ultra-pure gases (concentration of impurities < $10^{-5}$ %) the higher the $E_i$ value, the lower is the maximum achievable gain.

In hole-multipliers, e.g. GEMs and THGEMs, the avalanche is mostly confined within the holes; the cathode is screened from the direct light emitted by the avalanches (particularly in cascaded structures) and only a small fraction of the photons have a chance to impinge on it. Thus, the coefficient b << 1 (formula 2) and the main contribution to feedback is expected to originate from ions. Independently, the general trend should remain: the lowest achievable gain should be in gases having the highest ionization potentials. This tendency was confirmed in [18] where it was observed that in ultra-pure He and Ne a GEM could reach only gains close to unity.

In the present work we observed, however, that the highest gain was reached with a single-THGEM operating in Ne rather that in Ar (although the level of impurities in Ar was higher). The obvious explanation is that we did not reach yet ultra-pure conditions in our chamber (as in [18]); minute impurities in Ne were found to affect in a stronger way the values of $\gamma_{ph}$ and $\gamma_+$ than in Ar. These impurities have double effect: they change the emission spectra of the avalanches S(V, Ev) and via charge-exchange mechanism they modify the ion species impinging on the cathode. In the following we will discuses these effects separately.

**4.1. Emission spectra in the presence of impurities**

As was mentioned in the previous section, in clean noble gases avalanches emit eximer spectra [34, 35]. However, in the presence of small impurities dramatic changes in the emission spectra may occur [36, 37]. As was shown in [38] in relatively weak electric fields ( < 10 kV/cm) and in the presence of small concentrations of molecular additives in noble gases, only the spectra of the molecular impurities are excited. In very intense electric fields (~ $10^5$ V/cm) both the molecular spectrum and that of the noble gas are excited [37]. Note that even in this case the number of exited noble-gas atoms produced in the avalanche will be much higher that the number of their ions. Under moderate fields (typical to THGEM holes) the higher is the value $\Delta E = E_{in} - E_{imp}$ (where $E_{in}$ and $E_{imp}$ are the ionization potentials of the noble gas and the given molecular impurity, respectively), the lower is the probability to excite the noble gas [38]. As calculated in [26] the typical value of the electric field within the THGEM holes in Ar is below 20 kV/cm. Since in Ar $\Delta E$ is small and in Xe it is even smaller or in case of some impurities, for example $N_2$, it is even $\Delta E < 0$, one can expect that in pure Ar and Xe (recall that in our case the concentration of impurities was: $N_2$ ~0.01% and $H_2O$ ~0.7 % in Ar and $N_2$ ~ 0.003 %, and "$CF_4$ equivalent" impurities[*] ~0.06% in Xe), excimer continua are surely exited. Even though we did not perform any spectral analysis of the emitted photons, the excimer emission spectra of Ar were observed in similar conditions by others (see for example [36, 39]). However, in the case of Ne the situation is less clear, because $\Delta E$ is large and the electric field within the THGEM holes is almost 3 times lower than in Ar. It is most probable that under such relatively weak electric fields the Ne continuum was not excited due to the presence of the small amount of impurities (see explanations given in [38]). An additional factor is the large cross section (~ $10^{-14}$ $cm^2$ [40]) of the resonance energy transfer from excited Ne atoms to $N_2$ ionization.

**4.2 Charge exchanges in noble-gas ions with molecular impurities**

Because in Ne containing some impurities $\Delta E >> 0$, a great fraction of ions produced in the avalanche will be that of the impurities [38]. Moreover, even if some Ne ions and some Ne excited atoms are produced in the avalanche, they will effectively transfer their energy to molecules of the impurities; the main channel is through ionization. Thus during their drift to the cathode, Ne ions have high probability to get neutralized (even at small concentration of impurities); ions of impurities' molecules will then continue drifting to the cathode. In this scenario, at some critical concentration of impurities only their ions will reach the THGEM cathode. Due to their lower ionization potential and molecular structure (the coefficient $k_g$ in formula (5) for molecular impurities, $k_{gim}$, is smaller than that of Ne [41]) the probability to extract electrons from the cathode is considerably lower compared to that of Ne ions. This dramatically reduces ion feedback, increasing the maximum achievable gain, as was clearly observed in the present work - without the getter. It is hard to estimate the exact concentration of impurities at which this effect occurs, from simple considerations; the effect can be simulated, taking into account ion-transport parameters and charge-transfer processes. However, our results suggest that in the case of Ne, only a fraction of Ne ions ($Ne^+$) reached the cathode; others transferred their charge to the impurity molecules, which reduced the effective $\gamma_+$.

---

[*] The Residual Gas Analyzer could not resolve some of the impurities, identifying them as "$CF_4$".

In contrast, in Ar containing some impurities, due to the smaller value of $\Delta E$, more Ar ions will be produced in the avalanche (compared to Ne); thus a larger fraction of $Ar^+$ will reach the cathode - inducing ion feedback.

In Xe, which has relatively low $E_i$ value, $\gamma_+$ is consequently low; this resulted in only slightly reduced gain following gas purification (note that the different orientation of the impinging alpha particles did not affect the maximum gain). The effect of impurities in Xe is much less pronounced than in Ne or Ar due to small or even negative value of $\Delta E$.

### 4.3 Importance of dopants in noble-liquid detectors

As mentioned earlier, there have been efforts by several groups to implement hole-type gaseous multipliers (GEMs and THGEMs) in cryogenic dark-matter detectors (see for example [3, 20]). The present results indicate that in these conditions both multipliers may have limited gain, calling for either cascaded-multipliers approach or for doping noble liquids with impurities. As an example, in [42], Ar-$N_2$ (liquid Ar doped with $N_2$) was studied in a two-phase double-GEM detector; a 2-fold increase of gain (400 instead of 200) was obtained by adding 1.5% $N_2$ to Ar at T=84K, at a pressure 0.70 atm. The higher gain in nitrogen-doped Ar gas phase was, however, obtained at the expense of reduced primary scintillation in the liquid phase. This would be of a serious drawback in applications of two-phase detectors, where primary scintillation in the liquid is essential for background rejection. The results of our additional relevant studies of the effect of $N_2$ additives to Ne on the maximum gain, in THGEM, are described in Appendix A. Our work and that of [43] indicate that additional studies should be done to carefully optimize the nature of dopants and their concentration in noble-liquid rare-events detectors.

### 5. Summary

In this work, we presented results on the influence of impurities, particularly nitrogen, on the operation of a THGEM detector in Ar, Ne and Xe. Nitrogen appears to be playing a major role in allowing for high gains in these gases, most probably due to ion charge exchange. We estimate that $H_2O$ concentration did not play a major role; it was always rather high and remained as such even after getter purification. A clear correlation was observed in this work between the $N_2$ levels and maximum attainable gains. In both Ne and Ar, the gain dropped significantly when $N_2$ concentrations reached values below 0.2 %.

In purified Ne, the maximum gain $A_{mTHGEM}$ measured with alpha-particles dropped in a single-THGEM by a factor of ~20 ($N_2$ ~ 0.003%, $H_2O$ ~ 0.05%) and in double-THGEM by a factor of ~30 ($N_2$ ~ 0.002%, $H_2O$ ~0.005%). In purified Argon ($N_2$ ~ 0.01 %, $H_2O$ ~ 0.7 %), $A_{mTHGEM}$ dropped by a factor 30, yielding a value of 3 in a single-THGEM. The shift observed in operation voltage (for a given gain), could be indicative of a Penning effect in Ne-$N_2$ (see Appendix A); such effect was not observed in Ar-$N_2$.

The experimental data allow assuming that in our conditions the eximer continuum of Ne was not excited and only a fraction of ions reaching the cathode were $Ne^+$. In Ar, though less clean than Ne, the eximer continuum was most probably excited; in addition, a larger fraction of $Ar^+$ might have reached the cathode. As a result, in our present experimental conditions the maximum achievable gain in Ar (with some impurities) was somewhat lower than in higher-purity Ne.

The observed effects related to impurities are relevant to applications of noble-liquid detectors; of main concern are two-phase detectors with gas-avalanche readout in the gas phase, projected for rare-event experiments (e.g. Dark Matter, Solar Neutrino etc.) [3]. The origin of the low gain (~10) in ArDM two-phase LAr-TPC [43]) could by related to the lack of impurities; gain reduction due to freezing of impurities at low temperatures was also reported in [18]. The results may indicate upon an advantage of ionization-induced secondary-scintillation photon recording with gaseous photomultipliers [44] operating in quenched gases rather than in the noble-gas vapor. Further studies are required to assess the role of $N_2$ and other impurities in the gas phase of cryogenic detectors. Additional information on the role of impurities in noble-gas avalanches would require careful analysis of the avalanche emission spectra.

**Appendix: Study of the effect of $N_2$ additives on the maximum gain**

In an additional series of measurements aimed at assessing the influence of small $N_2$ additives to Ne, the gain was measured in a gas *flow mode* with a double-THGEM irradiated by a $^{55}$Fe X-ray source. $N_2$ content in Ne was regulated with the mass-flow controller in the range of 0.25 to 1 %; below 0.25 %, $N_2$ content was measured by the residual gas analyzer. Oxygen and water contents, originating primarily from the chamber walls, were on the order of $10^{-3}$ % their values remained fairly constant throughout the measurements.

The gain curves in Ne-$N_2$ are shown in Figure 6. With $N_2$ levels > 0.3 %, the double-THGEM gain reached values as high as $10^5$; they started dropping for $N_2$ values < 0.25 %. Note that the gain curves shifted towards the lower-voltage direction with $N_2$ decrease. In addition to deliberate $N_2$ admixture, in the *flow mode*, experiments were carried out with Ne flowing through the chamber after shutting off the $N_2$ gas line; the residual $N_2$ level gradually decreased with the gas flow, resulting in additional gain drop (Figure 6). Part of the remaining residual $N_2$ was due to air leaks and materials out-gassing (with no attempts made here to identify their origin).

At the lowest level of residual $N_2$ (~0.02 % in Ne), the maximum gain was about 2000; this value is about 100 times lower compared to the gain level recorded with Ne-0.5%$N_2$ in *flow mode*. Note that at the very low $N_2$ levels (without deliberate $N_2$ supply), in contrary to the results under $N_2$ flow, the gain curves shifted towards higher voltages.

Our measurements were concentrated on the effect of nitrogen only; the roles of other low-level impurities, particularly those falling below the RGA's detection limit (e.g. $O_2$) were not assessed; however, since the decrease in $N_2$ coincided well with the drastic gain reduction, it is expected to be the main cause for the observed change.

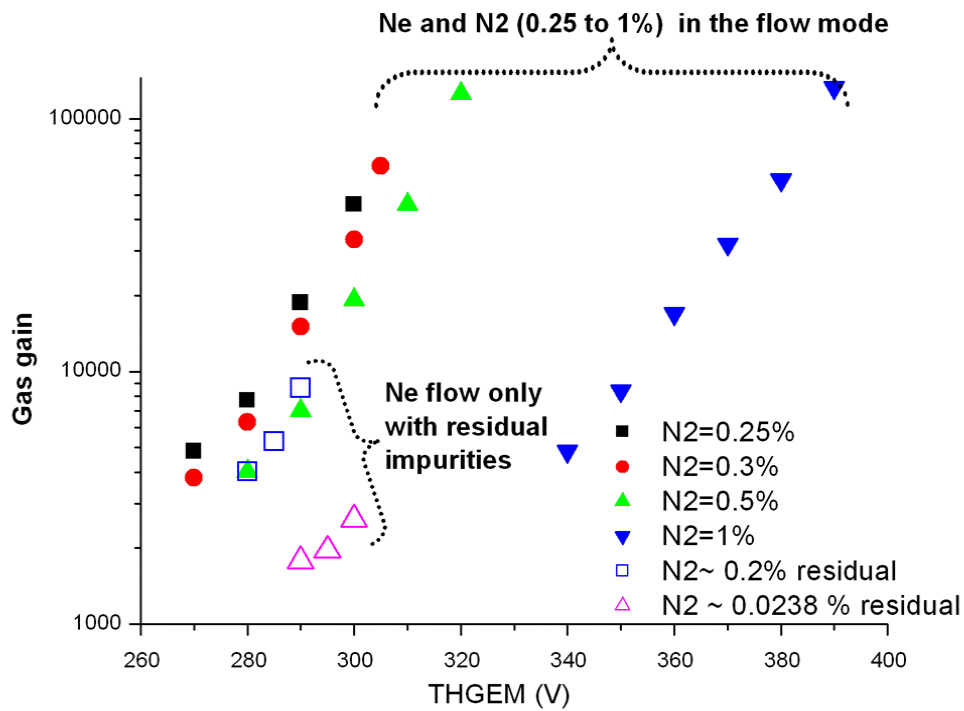

*Figure 6. Gain curves measured with the double-THGEM of Fig. 1 in atmospheric Ne-N$_2$ mixtures, measured with 6 keV x-rays. Measurements were carried out in gas flow mode, without purification, with Ne-N$_2$ mixtures (0.25 to 1 %) and with pure Ne flow at gradually decreasing residual N$_2$.*

## Acknowledgments

The work was partially supported by the Israel Science Foundation, grant No. 402/05 and by the Benozyio Center for Experimental Physics. J.M. was partially supported by the Israel Ministry of Integration; V.P. acknowledges the support of the Weizmann Institute of Science. Technical support by Mr. M. Klin of the Weizmann Institute and efficient cooperation of Print Electronics Ltd. are greatly acknowledged. A. B is the W.P. Reuther Professor of Research in The Peaceful Use of Atomic Energy.